\documentclass[aps,preprint,amssymb]{revtex4}

\usepackage{amsmath}

\usepackage{epsfig}
\usepackage{graphicx}
\usepackage{epstopdf}

\def\cm{cm$^{-1}\,$}
\def\h3o2{H$_{3}$O$_{2}^-$}
\def\HOHOH{[H--O$\cdots$H$\cdots$O--H]$^-\,$}
\def\HODOH{[H--O$\cdots$D$\cdots$O--H]$^-\,$}
\def\DOHOH{[D--O$\cdots$H$\cdots$O--H]$^-\,$}
\def\DOHOD{[D--O$\cdots$H$\cdots$O--D]$^-\,$}
\def\HODOD{[H--O$\cdots$D$\cdots$O--D]$^-\,$}
\def\DODOD{[D--O$\cdots$D$\cdots$O--D]$^-\,$}

\def\bR{\mbox{\boldmath $R$}}
\def\bP{\mbox{\boldmath $P$}}
\def\bQ{\mbox{\boldmath $Q$}}
\def\bg{\mbox{\boldmath $g$}}
\def\nl{\nonumber \\}
\def\re#1{(\ref{#1})}
\newcommand{\Fig}[1]{Fig.\,\ref{#1}}
\begin{document}
\title{A full-dimensional quantum dynamical study of the vibrational ground state of \h3o2 and its isotopomers}
\author{Yonggang Yang, Oliver K\"uhn}
\email[]{E-mail: oliver.kuehn@uni-rostock.de}
\thanks{New address: Institut f\"ur Physik, Universit\"at Rostock, Universit\"atsplatz 3, D-18051 Rostock}

\affiliation{Institut f\"ur Chemie und Biochemie, Freie Universit\"at Berlin, Takustr. 3, D-14195 Berlin}

\date{\today}

\begin{abstract}
\noindent We investigated the effect of deuteration on the vibrational ground state of the hydrated hydroxide anion using a nine-dimensional quantum dynamical model for the case of $J=0$. The propagation of the nuclear wave function has been performed with the multi-configuration time-dependent Hartree method which yielded zero-point energies for the normal and fully deuterated species in quantitative agreement with previous diffusion Monte Carlo calculations. According to the zero-point energy the isotopomers having the hydrogen atom in the bridging position are more stable by about 1 kJ/mol as compared to the deuterium case. This holds irrespective of the deuteration state of the two OH groups.  We also report the secondary geometric H/D isotope effect on the O--O distance which amounts to an elongation of about 0.005 \AA{} for the symmetric isotopomers and 0.009 \AA{} in the asymmetric case. Finally, we explore the isotopomer sensitivity of the ground state tunneling splitting due to the torsional motion of the two OH groups.
\end{abstract}

\maketitle
\section{Introduction}
\label{sec:intro}
Charged clusters with strong hydrogen-bonds have attracted considerable interest recently due to the progress of both, infrared spectroscopy and multidimensional quantum dynamical modeling \cite{asmis07:53,roscioli07:249}. In particular the hydrated proton and its negative analogue, the hydrated hydroxide anion, have been investigated in quite some detail. For example, the Zundel cation, H$_{5}$O$_{2}^+$,
became the paradigm for the predictive power of ab initio quantum dynamical calculations \cite{vendrell07:6918,vendrell07:184302,vendrell07:184303}.
The monohydrated hydroxide anion, \h3o2, has been studied by Johnson and coworkers using argon predissociation spectroscopy.  In the range above 3000 \cm a sharp doublet has been detected \cite{price02:412,robertson03:1367} at about 3650 \cm which was subsequently assigned to emerge from transitions between the HOOH torsion doublet and the ``free'' OH stretchings  \cite{huang04:5042}.
The spectral range from 1000-1900 \cm  is dominated by a peak at 1090 \cm which was attributed to result from a combination of bridging hydrogen (BH) stretch, wag, and rock motions of the whole complex \cite{diken05:571}. Effectively, this corresponds to a displacement of the BH away from the O-O axis. The region below 1000 \cm was finally addressed in Ref. \cite{diken05:1487}. Besides two smaller features at 940 \cm and 995 \cm, with one being the second perpendicular mode of the BH, the spectrum is dominated by an intense and rather narrow peak at 697 \cm which has been assigned to the BH motion along the O-O axis.

The principal difficulty in assigning the \h3o2 spectrum arises from the fact that similar to the Zundel cation one has to deal with a strong hydrogen bond in a floppy structure. In other words, the vibrational dynamics is rather anharmonic and any calculation bound to the harmonic approximation is likely to fail at least for the BH modes. Moreover, this system features a double minimum potential with a barrier of only about 75 \cm \cite{samson02:201}. Therefore, the classical equilibrium structure will be nonsymmetric, e.g. [H--O--H$\cdots$O--H]$^-$, but as the zero-point energy (ZPE) including the BH motion is above the barrier the symmetric C$_{2}$ structure [H--O$\cdots$H$\cdots$O--H]$^-$ will be stabilized \cite{samson02:201,xantheas95:10373,tuckerman97:817} as in other hydrogen-bonded clusters \cite{asmis07:8691}. Of course, this affects especially the transition frequency of the BH stretching vibration drastically. Klopper et al. \cite{samson02:201} calculated a one--dimensional potential energy curve for this vibration which gave a ZPE about 350 \cm above the barrier and a fundamental transition frequency of about 1030 \cm which is lower than the harmonic prediction for a nonsymmetric structure (BH stretch coupled to the water bend around 1600-1700 \cm \cite{xantheas95:10373}), but still far above the experimental value of 697 \cm \cite{diken05:1487}.  This points to the need for a multidimensional treatment of the infrared spectrum as reported  by J. Bowman and coworkers who used multimode reaction path  vibrationally self-consistent field/configuration interaction (VSCF/CI) and diffusion Monte Carlo (DMC) techniques \cite{huang04:5042,mccoy05:064317}. For the specific case of the BH stretch these methods gave 741 \cm and 644 \cm, respectively. The accuracy (most modes within 38 \cm) of the VSCF/CI was recently confirmed by Lanczos diagonalization calculations in Ref. \cite{yu06:204306},

In this contribution we will focus on the effect of H/D isotopic substitution on the properties of the hydrogen-bond in this complex. Our primary goal is to answer the question whether deuteration of the hydroxide ([D--O$\cdots$ H$\cdots$ O--H]$^-$) or the ``solvent'' water ([H--O$\cdots$D$\cdots$ O--H]$^-$ yields an energetically  more stable structure. This issue is related to the problem of isotopic exchange equilibriums and associated fractionation factors which have been discussed for gas phase reactions of hydrogen--bonded ions \cite{weil85:6859,larson88:1087,diken04:17} as well as in aqueous \cite{more-oferral80:1840,chiang80:1832} and other \cite{kreevoy77:5207,kreevoy80:3315} solutions. An extensive account on the stability of various charge and neutral water clusters has been given by Scheiner and coworker employing the MP2 method together with a (small) 6-31+G** basis set \cite{scheiner96:1511}. The picture which emerged upon comparing harmonic ZPEs has been as follows: In neutral clusters like the water dimer the O$\cdots$D$\cdots$O hydrogen bond is stronger than the O$\cdots$H$\cdots$O one mostly due to the intermolecular vibration of the bridging proton/deuteron. In charged clusters O$\cdots$H$\cdots$O wins as a consequence of the change in ZPE of the intramolecular water OH/D stretching vibration. 
In passing we note that for the Zundel cation this conclusion has been recently confirmed using anharmonic  calculations\cite{devlin05:439,mccunn08:321}.
For \h3o2 the difference in ZPE is as small as about 50 \cm, although it was argued that this effect is enhanced by entropic effects. A related question could be asked for the doubly substituted case. Here, harmonic analysis predicts \DOHOD to be slightly favored over \HODOD.
It is important to emphasize that the conclusions for \h3o2 have been drawn from harmonic calculations with respect to  nonsymmetric equilibrium structure.
In view of the discussion above this calls for having a second look at this problem from the perspective of a quantum mechanical treatment in full dimensionality.

The change in strength of the hydrogen bond upon isotopic substitution is also reflected in geometric isotope effects (GIE) which is yet another manifestation of the multidimensional anharmonic nature of the  potential energy surface (PES) \cite{giese06:211,shibl07:315}. For weak hydrogen bonds, having a double minimum PES,  H/D substitution leads to a shorter O--D distance as compared to O--H. This in turn weakens the hydrogen bond thus causing the O--O distance to increase. For strong symmetric hydrogen bonds where the ZPE is above the barrier and the vibrational distribution has its maximum at the barrier top, H/D substitution reduces the width of this distribution which pulls the oxygens towards the deuterium, that is, the  O--O distance decreases.  In the present paper we will address the GIE for all isotopomers and different degrees of deuteration on the basis of the full-dimensional ground state wave function, that is, at zero temperature.

In the following Section \ref{sec:hamil} we will present a nine-dimensional (9D) Hamiltonian which describes the vibrational motion of \h3o2 in full-dimensionality for total angular momentum equal to zero. This Hamiltonian is based on the PES developed by Bowman and coworkers \cite{mccoy05:064317}. The related operator for the kinetic energy is given in the Appendix. The vibrational ground state is obtained by imaginary time-propagation using the Multi-configuration Time-dependent Hartree (MCTDH) method \cite{beck00:1,meyer03:251}; numerical details of the calculation are given in Section \ref{sec:num}. In Section \ref{sec:results} we will present results on the vibrational ground state of the different isotopomers and on the GIE. The paper is summarized in Section \ref{sec:sum}.
%
\section{Theoretical Model}
\label{sec:theory}
\subsection{9D Hamiltonian}
\label{sec:hamil}
\begin{figure}[t]
\begin{center}
\centerline{\includegraphics [scale=0.6] {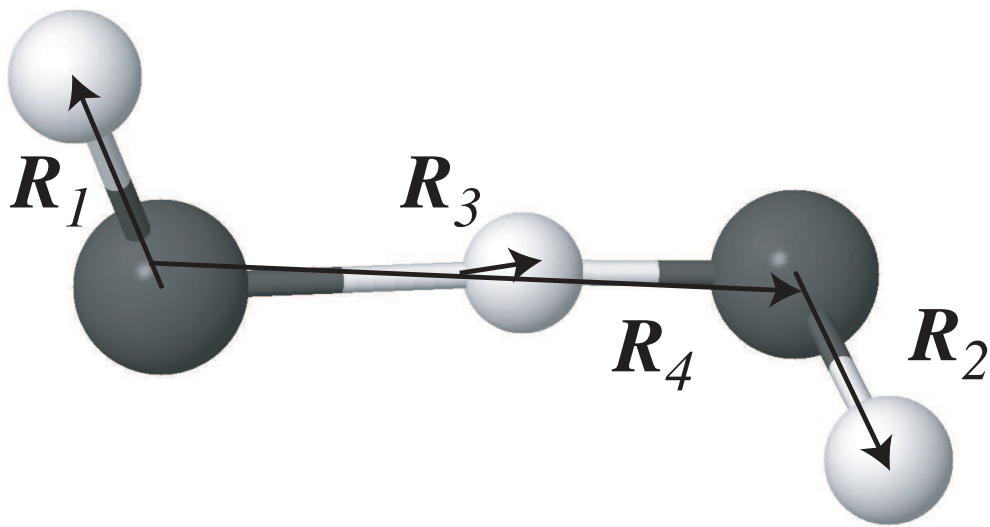}}
\end{center}
\caption{Classical equilibrium structure of the \h3o2 anion and the four Jacobi vectors used for defining the nine internal coordinates (see text). For the PES expansion in Eq. \re{eq:pes} we have used a reference geometry based on the coordinate expectation values as obtained for \h3o2 (cf. Tab. \ref{tab:mctdh}). 
}
\label{fig:coords}
\end{figure}

The choice of coordinates is crucial as it determines the strength of correlations between different degrees of freedom in the potential and kinetic energy operator. For the present case the issue is complicated as there is the possibility of large amplitude torsional motion. This made it necessary to use a reaction path approach in Ref. \cite{mccoy05:064317}, which combined the torsional reaction coordinate with orthogonal normal mode displacements taken with respect  to the C$_{2}$ transition structure. Here we will use internal coordinates, which is less restrictive but comes at the expense of a more complicated kinetic energy operator. After separating the total center of mass, the four Jacobi vectors shown in Fig. \ref{fig:coords} will be used: $\bR_1$ and $\bR_2$ each connecting one oxygen and the ``free'' hydrogen atom, $\bR_4$ connecting two centers of mass of the OH groups, and  $\bR_3$ connecting the shared hydrogen atom and the center of mass of the  O$_2$H$_2$ fragment (accordingly for the deuterated cases). Assuming the total angular momentum $J^2=0$ the kinetic energy operator  depends on the internal coordinates only. Based on these Jacobi coordinates the following nine internal coordinates are  chosen: the lengths $R_1$, $R_2$, and $R_4$, of the vectors $\bR_1$, $\bR_2$, and $\bR_4$, respectively, the three Cartesian components of the BH, $x$, $y$, and $z$, being the components of $\bR_3$, the angle $\theta_1$  ($\theta_2$) between $\bR_1$ and $\bR_4$ ($\bR_2$ and $\bR_4$), and the dihedral angle $\phi$ between the planes spanned by the vectors ($\bR_1$, $\bR_4$) and ($\bR_2$, $\bR_4$) which describes the torsional motion. Notice that due to its definition with respect to the Jacobi vectors, $\phi$ is slightly different from the torsional coordinate used in Ref. \cite{mccoy05:064317}. For the origin we use the center of mass and $z$ is defined along the direction of $\bR_4$, i.e. it roughly corresponds to the BH stretch mode.

In principle the exact kinetic energy operator can be obtained in terms of arbitrary coordinates by proper quantization of the classical Lagrangian. As it contains hundreds of terms one needs to compromise and we neglected the coupling between the angular momenta of the BH and the  O$_2$H$_2$ fragment. For $J^2=0$ this yields the kinetic energy operator given in the Appendix which contains 21 terms only. Notice that in the following numerical simulations we will use the new variables $u_i=\cos \theta_i (i=1,2)$.

The full-dimensional potential energy surface is constructed by cumulative expansion of different correlation orders \cite{carter97:10458}. Following the strategy of Ref. \cite{vendrell07:184302} we will combine certain groups of coordinates to treat their correlation exactly. Specifically we have chosen the three groups $\bg_1=[R_1, R_2, R_4]$, $\bg_2=[u_1, u_2, \phi]$, and $\bg_3=[x, y, z]$. The final expansion we used to generate the 9D PES reads
\begin{equation}
\label{eq:pes}
V(\bg_1,\bg_2,\bg_3)=V_{0}+ \sum_{i}V^{(1)}(\bg_i)+ \sum_{i<j}V^{(2)}(\bg_i,\bg_j),
\end{equation}
where $V^{(n)}$ gives the  $n$-mode correlation between sets of coordinates. Notice, however, that this PES contains up to 6-mode correlations between individual coordinates. In Eq. \re{eq:pes} $V_0$ is the energy of the reference geometry, $\bg^{(0)}$. An obvious choice for this reference is the classical transition state geometry. We have performed a ground state calculation based on this reference and obtained an error of about 2-3 \cm per degree of freedom as compared to the Quantum Monte Carlo results for \h3o2 reported in Ref. \cite{mccoy05:064317}. In a next step we have attempted to improve  this result and found that
one can get essentially a quantitative agreement if the quantum mechanical ground state expectation values (as obtained from the classical reference calculation) for  the \h3o2 coordinates are used to define the reference. For simplicity we have employed this reference for all isotopomers (see Table \ref{tab:mctdh}).

For generating the PES in the given coordinates we have used the fitted CCSD(T)/aug-cc-pVTZ potential of Bowman and coworkers \cite{mccoy05:064317}. Two representative cuts of the PES are shown in \Fig{fig:pes}.

\subsection{MCTDH Propagation}
\label{sec:num}
The vibrational ground state of the different isotopomers has been obtained from numerical solution of the time-dependent Schr\"odinger equation in imaginary time \cite{kosloff86:223}. For the propagation we used the MCTDH method \cite{beck00:1,meyer03:251} as implemented in the Heidelberg program package \cite{mctdh84}. Here, the wavefunction is expanded as a superposition of Hartree-products consisting of time-dependent single particle functions (SPFs). A prerequisite for efficient propagation is a Hamiltonian which has product form. While the kinetic energy operator fulfills this requirement, the potential has been fitted to a sum of products using the POTFIT approach \cite{meyer03:251}.  For the representation of the SPFs a discrete variable representation (DVR) has been utilized. For the torsional coordinate $\phi$ an
exponential DVR representation with periodic basis functions (eigenfunctions of $\frac{d}{d \phi}$) was used. All other coordinates have been expressed via a harmonic oscillator DVR. The SPF basis functions cover
the  range energies below 10000 \cm; the smallest natural orbital population in the ground state was 0.0003. All parameters are compiled in Tab. \ref{tab:mctdh}.

\section{Results}
\label{sec:results}
\subsection{Vibrational Ground State}
\label{sec:GS}
\begin{figure}[t]
\begin{center}
\includegraphics[scale=0.6]{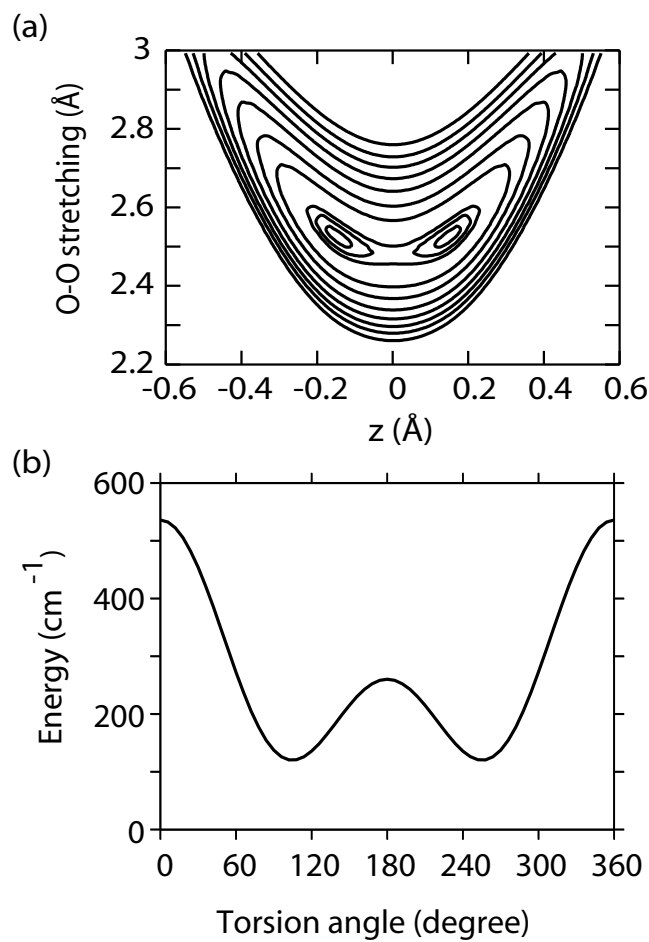}
\end{center}
\caption{(a) PES along O-O distance ($R_{4}$) and proton transfer coordinate $z$ (contour lines at (in \cm): 230, 260, 300, 600, 900, 1300,  1700, 2100, 2500, 3000). (b) Potential energy curve along the coordinate $\phi$ which for the symmetric case corresponds to the usual torsion angle. In both panels all other coordinates have been kept frozen at the transition state geometry.}
\label{fig:pes}
\end{figure}

\begin{figure}[t]
\begin{center}
\includegraphics[scale=0.6]{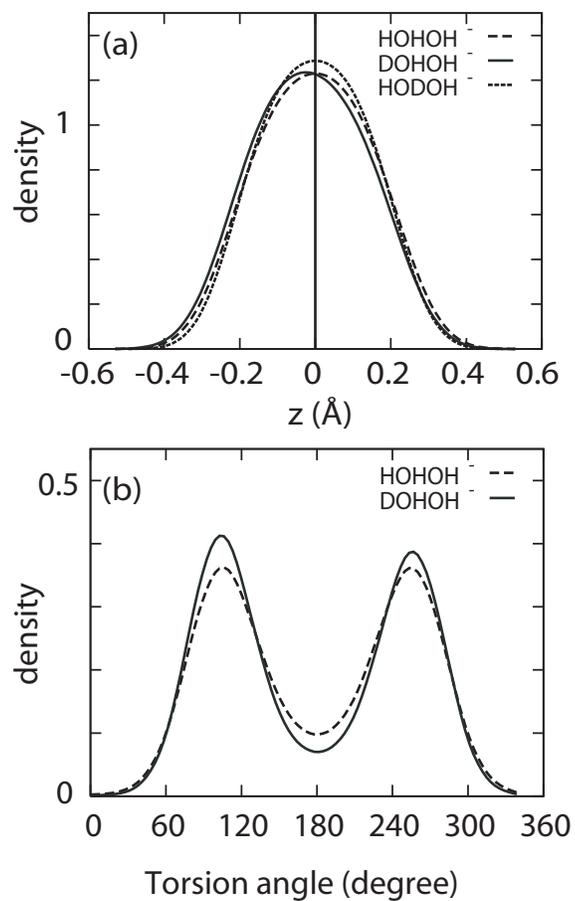}
\end{center}
\caption{\label{fig:psi} Reduced probability density for the vibrational ground states of the different isotopomers for the proton transfer (a) and the torsional coordinate (b).}
\end{figure}

Representative cuts of the full 9D ground state vibrational density are shown for the different isotopomers in Fig. \ref{fig:psi} and the coordinate expectation values and their variances are compiled in Tabs. \ref{tab:coords} and \ref{tab:coords2}. Fig. \ref{fig:psi}a gives the densities along the shared proton coordinate $z$. For the symmetric cases \HOHOH and \HODOH the distributions have their maxima at $z=0$ indicating the symmetrization due to ZPE which is above the classical barrier in Fig. \ref{fig:pes}a. As expected the width of the distribution narrows upon isotopic substitution. Inspecting Tab. \ref{tab:coords} one finds, that this amounts to $\sim$6\% for the $z$ coordinate, but to as much as 13 \% 
for the bridging hydrogen's  bending coordinates $x$ and $y$. The distributions along the other coordinates are much less affected, e.g., the difference for the torsion coordinate $\phi$ is not noticeable on the scale of Fig. \ref{fig:psi}b. Asymmetric isotopic substitution in \DOHOH introduces changes mostly at the substitution site, that is, $R_{1}$ decreases and the associated width drops by 15\%. 
Similar changes in the widths are observed for the angles $u_{1}$ and $\phi$ while the widths for the BH coordinates are essentially unaffected. The introduced asymmetry can also be deduced from the distribution along the  coordinate $\phi$ in Fig. \ref{fig:pes}b. 
The widths for the bridging hydrogen do not change much in \DOHOH, however, as compared to \HOHOH the structure becomes asymmetric also insofar as the hydrogen atom moves toward the OH group and out of the O--O axis, i.e. [D--O$\cdots$H--O--H]$^-\,$ is formed.  
Related trends are observed if one starts from \DODOD and makes D/H substitution. The definition of the coordinates depends on the isotopomer which affects the asymmetric cases. Therefore,  we also give the values of internal coordinates in the table captions in order to support the conclusion mentioned above. 

Next we focus on the ZPEs of the different isotopomers which are analyzed in terms of the contributions of the different parts of the Hamiltonian in Tab. \ref{tab:zpe}.  First we notice, that the ZPE of \HOHOH is calculated as 6606 \cm which essentially reproduces the DMC result (6605$\pm$ 5 \cm) obtained on the \emph{fully coupled} PES \cite{mccoy05:064317}. The set $\bg_{1}$ gives the largest contribution to the ZPE as it contains the high frequency OH stretching vibrations. Further, the strongest correlations between different sets of coordinates are those involving the BH motion. For the fully deuterated case we obtain 4481 \cm which is also in accord with the DMC result (4487$\pm$ 5 \cm).

Next we discuss the general trend in ZPE change upon H/D substitution. Inspecting Table \ref{tab:zpe} we observe that replacing H by D in one of the OH groups lowers the ZPE by about 600 \cm irrespective whether the other OH group is deuterated or not. Replacing H by D in the bridging site lowers the total ZPE by about 520 \cm only, again irrespective of the OH groups.  In other words, in terms of the ZPE the H-bond is about 80 \cm stronger than the D-bond. Let us have a more detailed look at the single substitution case. Here, we find that in terms of ZPE \DOHOH is more stable than \HODOH by 82 \cm. First of all, this confirms the qualitative result of Scheiner and coworker \cite{scheiner96:1511} who obtained a value of 52 \cm based on the harmonic approximation of the PES around a nonsymmetric structure. Their normal mode treatment led to the conclusion that it is the intramolecular water OH(D) stretching which is responsible for the  increased stability of \DOHOH. However, as discussed before the structure is symmetrized by ZPE and its analysis in terms of a water molecule hydrogen-bonded to a OH(D)$^-$ is not adequate. Instead inspecting Tab. \ref{tab:zpe} we find the following behavior: First, ZPE changes in \DOHOH are due to the set $\bg_{1}$, i.e. mostly the OH-stretching at the deuteration site, while in \HODOH it is set $\bg_{3}$ involving the BH. 
We also notice that the net effect of all single mode potentials is 37 \cm by which the D-bond would be more stable whereas the correlation energy difference of -119 \cm finally leads to the preference for the H-bond. Although we should emphasize that the actual values of the different energy contributions depend, of course, on the chosen reference geometry for the PES expansion, it seems plausible that, for instance, the correlation between the bridging hydrogen set $\bg_{3}$ and both sets $\bg_{1}$ and $\bg_{2}$ plays a significant role for the ground state wave function. As a caveat we note that these findings do not necessarily imply a simultaneous coupling between all modes of the given sets.

%
\subsection{Secondary geometric isotope effect}
%
Having at hand the ground state wave functions we can calculate
the secondary GIE, i.e., the change of O--O distance upon isotopic
substitution. Since $R_{4}$ does not fully correspond to the O--O
distance, $R_{\rm O-O}$, (cf. Fig. \ref{fig:coords}) the
respective operator had to be expressed in terms of  our model
coordinates.  The resulting expectation values for the different
isotopomers are compiled in Tab. \ref{tab:oo}. The classical value
for $R_{\rm O-O}$ in HOHOH$^-$  at the MP2/aug-cc-pVTZ transition
state is 2.446 \AA. In the quantum case this value increases to
2.492 \AA{} as a consequence of zero-point vibration. This value
is reduced to 2.488 \AA{} for substitution of the BH, i.e. in the
HODOH$^-$ case. In all cases the substitution of the central BH
leads to a reduction of the O--O distance. For the symmetric cases the
reduction is about 0.005 \AA{} while for the asymmetric case it is 0.009 \AA.
Such a bond compression due to
reduced zero-point vibration, i.e., localization of the wave function, is typical for strong hydrogen
bonds.

The changes reported here are in good agreement with the DMC calculations
in Ref. \cite{mccoy05:064317} for H$_{3}$O$_{2}^-$ and D$_{3}$O$_{2}^-$, only the absolute values of the DMC  $R_{\rm O-O}$ are larger by 0.005 \AA.
In passing we note that the fully deuterated case
has also been investigated using the finite temperature path integral method at a lower level of quantum chemistry
\cite{tachikawa05:11908}. For this situation Tachikawa and coworker obtained 2.498 \AA{} for  \h3o2
and 2.504 \AA{} for D$_{3}$O$_{2}^-$, that is, the opposite trend which has been explained by the
bimodal character of the calculated distribution for D$_{3}$O$_{2}^-$ at the given level of theory.

%
\section{Summary}
\label{sec:sum}
%
The monohydrated hydroxide anion is a prototype for a  strongly
hydrogen-bonded low-barrier system whose structure is symmetrized
by zero-point vibrational motion. An accurate theoretical
prediction of the associated ground state wave function requires
to treat the dynamical problem in full-dimensionality. Using the
CCSD(T) potential energy surface of Bowman and coworkers
\cite{mccoy05:064317} we have shown that the MCTDH approach for
wave packet propagation \cite{beck00:1} can meet this challenge.
Comparing  the vibrational ground state wavefunctions and ZPEs for different isotopomers the following conclusions could be drawn: First, in accord with the general view that for ions bridging donor and acceptor by a hydrogen atom is more favorable than bridging by a deuterium. The general trend for the considered systems is that in terms of ZPE the H-bonds are about 80 \cm  ($\sim$1 kJ/mol) more favorable than the D-bonds, irrespective the deuteration state of the OH groups. Specifically, we find  \DOHOH to be energetically more stable than \HODOH by 82 \cm. Although this seems merely to confirm the results of the harmonic analysis reported in Ref. \cite{scheiner96:1511}, the present full-dimensional treatment is providing  a more realistic physical picture by accounting not only for the symmetrization of the structure due to zero point motion but also for the anharmonicity of the potential energy surface. Viewed from the perspective of harmonic vibrations it is not only the loss of an intramolecular OH vibration of  the water molecule which reduces the ZPE more than the loss of the intermolecular bridging hydrogen vibration. In fact we find that correlations between the bridging H/D atom motion and especially the O-O stretching contribute significantly to the ZPE. 

The second result concerns the H/D isotope effect on the heavy atom O--O distance. Here, we find that as compared with the classical prediction, zero-point vibrational motion increases $R_{\rm O-O}$ in the H-bonded isotopomers. Due to the reduced wave function delocalization in the D-bonded forms $R_{\rm O-O}$ decreases again, although not below the classical value. 

%
\section*{Acknowledgment}
%
This work has been financially supported by the Deutsche Forschungsgemeinschaft (GK 788). The authors gratefully acknowledge stimulating discussions with Prof. H.-H. Limbach  who pointed us to this problem in the context of fractionation factors. We thank Prof. J. Bowman and Dr. X. Huang for providing the code for the CCSD(T) potential energy surface.
%
\section*{Appendix}
\label{sec:apd}
%
The kinetic energy operator is derived by direct quantization of the classical kinetic energy
$T_{\rm class}=\frac{1}{2} \dot {\bQ} ^{\dagger} \mbox{\boldmath $M$}  \dot {\bQ}$, where $\bQ$ is a vector including all generalized coordinates and $\dagger$ means Hermitian conjugate which is equivalent to transpose in the classical kinetic energy. The corresponding momentum operator is defined by $\bP=-i \hbar \partial/\partial \bQ$ and hence the quantum kinetic energy operator is $T=\frac{1}{2} {\bP} ^{\dagger} \mbox{\boldmath $M$}^{-1} {\bP}$. The general method to derive the Hermitian conjugate of the momentum will be detailed elsewhere. The classical kinetic energy in laboratory reference frame is obtained by conventional mechanics as a sum of two parts $T_{\rm class}=T_{\rm class}(\bR_3)+T_{\rm class}(\bR_1,\bR_2,\bR_4)$. For the latter part in the sum we have ignored the global rotation of the O$_2$H$_2$ fragments. In other words the angular momentum coupling between the central hydrogen and the O$_2$H$_2$ fragments has been neglected in the expression for the exact kinetic energy operator for total angular momentum $J^2=0$. For the  simulation an additional normalization transform was performed to slightly reduce the numerical effort.

This gives the following 9D kinetic energy operator:
\[
T=T_{1}+T_{2}+T_{3}
\]
with
\begin{eqnarray}
T_{1}&=&-\frac{\hbar ^2}{2 \mu_1} \frac{\partial ^2}{\partial R_1^2} -\frac{\hbar ^2}{2 \mu_2} \frac{\partial ^2}{\partial R_2^2} -\frac{\hbar ^2}{2 \mu_4} \frac{\partial ^2}{\partial R_4^2}\nonumber
\end{eqnarray}
\begin{eqnarray}
T_{2}&=&  -\left(\frac{1}{2 \mu_1 R_1^2}+\frac{1}{2 \mu_4 R_4^2}\right) \frac{\partial}{\partial u_1} (1-u_1^2) \frac{\partial}{\partial u_1}
-\left(\frac{1}{2 \mu_2 R_2^2}+\frac{1}{2 \mu_4 R_4^2}\right) \frac{\partial}{\partial u_2} (1-u_2^2) \frac{\partial}{\partial u_2} \nl
&&-\frac{1}{2 \mu_4 R_4^2}\left(\sqrt{1-u_1^2} \frac{\partial}{\partial u_1} \frac{\partial}{\partial u_2} \sqrt{1-u_2^2} + \frac{\partial}{\partial u_1} \sqrt{1-u_1^2} \sqrt{1-u_2^2} \frac{\partial}{\partial u_2} \right) \nl
&&-\sum_{i=1,2}\left(\frac{1}{2 \mu_i R_i^2} \frac{1}{1-u_i^2}+ \frac{1}{2 \mu_4 R_4^2} \frac{u_i^2}{1-u_i^2}\right) \frac{\partial ^2}{\partial \phi^2} \nl
&&+\frac{1}{ \mu_4 R_4^2} \frac{u_1}{\sqrt{1-u_1^2}} \frac{u_2}{\sqrt{1-u_2^2}} \frac{\partial }{\partial \phi} \cos \phi  \frac{\partial }{\partial \phi} \nl
&&-\frac{1}{2 \mu_4 R_4^2} \frac{u_2}{\sqrt{1-u_2^2}} \left(\frac{\partial }{\partial \phi} \sin \phi \sqrt{1-u_1^2} \frac{\partial }{\partial u_1}+ \frac{\partial }{\partial u_1} \sqrt{1-u_1^2}  \sin \phi \frac{\partial }{\partial \phi}\right)\nl
&&-\frac{1}{2 \mu_4 R_4^2} \frac{u_1}{\sqrt{1-u_1^2}} \left(\frac{\partial }{\partial \phi} \sin \phi \sqrt{1-u_2^2} \frac{\partial }{\partial u_2}+ \frac{\partial }{\partial u_2} \sqrt{1-u_2^2}  \sin \phi \frac{\partial }{\partial \phi}\right)\nonumber \, .
\end{eqnarray}
\begin{eqnarray}
T_{3}&=& -\frac{\hbar ^2}{2 \mu_3} \frac{\partial ^2}{\partial x^2}
-\frac{\hbar ^2}{2 \mu_3} \frac{\partial ^2}{\partial y^2}
-\frac{\hbar ^2}{2 \mu_3} \frac{\partial ^2}{\partial z^2} \nonumber
\end{eqnarray}
The non-Euclidean normalization according to the volume element is $d_\tau = dR_1 dR_2 dR_4 dx dy dz du_1 du_2 d \phi$.
The reduced masses are defined as follows: HOHOH$^-$ -- $\mu_1=\mu_2=m_{\rm H}m_{\rm O}/(m_{\rm H}+m_{\rm O})$, $\mu_3=2m_{\rm H}(m_{\rm H}+m_{\rm O})/(3m_{\rm H}+2m_{\rm O})$, $\mu_4=(m_{\rm H}+m_{\rm O})/2$; HODOH$^-$ -- $\mu_3=2m_{\rm D}(m_{\rm H}+m_{\rm O})/(m_{\rm D}+2m_{\rm H}+2m_{\rm O})$;  HOHOD$^-$ --
$\mu_1=m_{\rm  D} m_{\rm O}/(m_{\rm H}+m_{\rm O})$ and $\mu_3$ and $ \mu_4$ change correspondingly.
For DODOD$^-$ the corresponding  masses of HOHOH$^-$ are modified by replacing $m_{\rm H}$ by $m_{\rm D}$.
In the same way we get the masses for DOHOD$^-$ by exchanging $m_{\rm H}$ and $m_{\rm D}$ in HODOH$^-$, and similarly one can obtain
HODOD$^-$ from DOHOH$^-$.
%

\clearpage
\newpage

\begin{table}[t]
\begin{center}
\caption{\label{tab:mctdh} MCTDH parameters for the imaginary time propagation (lengths in \AA, $N_{\rm DVR}$: number of DVR points; $N_{\rm SPF}$: number of SPFs.) and reference geometry, $\bg^{(0)}$, for the PES expansion in Eq. \re{eq:pes} (X=H/D). Note that the
coordinates of the reference geometry are based on expectation values for the HOHOH$^-$ case. The actual different values reported here for the different isotopomers are due to the fact that the Jacobi vectors defining the coordinate system are mass-dependent.
The classical transition state (TS) reference for HOXOH$^-$ is given as well.}
\begin{tabular}{c|ccc|ccc|ccc}
\hline
\hline
    &   $R_1$   &   $R_2$   &   $R_4$   &   $u_1$   &   $u_2$   &   $\phi$  &   $x$     &   $y$         &   $z$ \\
\hline
min. grid   &   0.740   &   0.740   &   2.275   &   -1.0        &   -0.7        &   0       &   -0.635  &   -0.635  &   -0.529 \\
max. grid   &   1.322   &   1.322   &   3.069   &   0.7         &   1.0         &   $2\pi$  &   0.635   &   0.635   &   0.529 \\
$N_{\rm DVR}$ & 11  &   11      &   16      &   13      &   13      &   17      &   13      &   13      &   17  \\
\hline
$N_{\rm SPF}$ & \multicolumn{3}{c|}{5} & \multicolumn{3}{c|}{10} & \multicolumn{3}{c}{13}   \\
\hline
\hline
$\bg^{(0)}$: HOXOH$^-$  &   0.979   &   0.979   &   2.519   &   -0.25         &   0.25         &   $\pi$  &   0.0   &   0.0   &   0.0 \\
$\bg^{(0)}$: DOXOD$^-$  &   0.979   &   0.979   &   2.547   &   -0.288         &   0.288         &   $\pi$  &   0.0   &   0.0   &   0.0 \\
$\bg^{(0)}$: DOXOH$^-$  &   0.979   &   0.979   &   2.532   &   -0.269         &   0.269         &   $\pi$  &   -0.012   &   0.021   &   0.043 \\
TS: HOXOH$^-$  &   0.962   &   0.962   &   2.477   &   -0.282         &   0.282         &   $\pi$  &   0.0   &   0.0   &   0.0 \\
\hline
\hline
\end{tabular}
\end{center}
\end{table}
\clearpage
\newpage
\begin{table}
\begin{center}
\caption{\label{tab:coords} Coordinate expectation values and their variances (in \AA) for the 9D ground states of \h3o2 and different single D substitution isotopomers.  The expectation values can be translated into internal coordinates which gives for DOH$^*$O$^*$H$^-$: $R_{\rm O-H^*}$=1.270 \AA, $R_{\rm O^*-H^*}$=1.226 \AA, angle$_{\rm O-H^*-O^*}$=178.8$^\circ$.}
\begin{tabular}{c|cc|cc|cc}
\hline
\hline
 & \multicolumn{2}{c} {HOHOH$^-$} & \multicolumn{2}{c} {HODOH$^-$} & \multicolumn{2}{c} {DOHOH$^-$} \\

 coordinate & mean & variance  & mean & variance & mean & variance\\
\hline
$R_{1}$     &   0.981   &   0.070   &       0.981   &   0.070   &       0.976   &   0.059\\
$R_{2}$     &   0.981   &   0.070   &       0.981   &   0.070   &   0.981   &       0.070\\
$R_{4}$     &   2.521   &   0.063   &       2.515   &   0.064   &       2.536   &   0.065\\
$u_{1}$     &   -0.259  &   0.173   &   -0.261      &   0.172   &   -0.267      &   0.149\\
$u_{2}$     &   0.259   &   0.173   &   0.261       &   0.172   &       0.269   &   0.174\\
$\phi$      &   3.138   &   1.240   &       3.136   &   1.238   &       2.886   &   1.258 \\
$x$     &   0.000       &   0.125   &   0.000   &   0.107   &   -0.022  &   0.126\\
$y$     &       0.000   &   0.120   &       0.000   &   0.104   &       0.029   &   0.123\\
$z$         &   0.000   &   0.150   &   0.000   &   0.140   &   0.029  &   0.152\\
\hline
\hline
\end{tabular}
\end{center}
\end{table}

\clearpage
\newpage
\begin{table}
\begin{center}
\caption{\label{tab:coords2} Coordinate expectation values and their variances (in \AA) for the 9D ground states of D$_3$O$_2^-$ and  different single H substitution isotopomers. The expectation values can be translated into internal coordinates which gives for DOD$^*$O$^*$H$^-$: $R_{\rm O-D^*}$=1.263 \AA, $R_{\rm O^*-D^*}$=1.225 \AA, angle$_{\rm O-D^*-O^*}$=178.9$^\circ$.}
\begin{tabular}{c|cc|cc|cc}
\hline
\hline
 & \multicolumn{2}{c} {DODOD$^-$} & \multicolumn{2}{c} {DOHOD$^-$} & \multicolumn{2}{c} {DODOH$^-$} \\
 coordinate & mean & variance  & mean & variance & mean & variance\\
 \hline
$R_{1}$         &       0.976  &       0.059 &       0.976   &       0.059   &       0.976   &       0.059\\
$R_{2}$         &       0.976  &       0.059 &       0.976   &       0.059   &       0.981   &       0.070\\
$R_{4}$         &       2.544  &       0.064 &       2.549   &       0.064   &       2.528   &       0.064\\
$u_{1}$         &       -0.281         &       0.148   &       -0.279  &       0.148   &       -0.270  &  0.146\\
$u_{2}$         &       -0.281         &       0.148   &       0.279   &       0.148   &       0.272   &  0.172\\
$\phi$          &       3.139          &       1.290   &       3.135   &       1.293   &       2.889   & 1.254 \\
$x$     &      0.000           &       0.120   &       0.000   &       0.135   &       -0.021 &       0.109\\
$y$     &       0.000           &       0.104   &       0.000    &       0.120   &       0.028  &       0.105\\
$z$     &       0.000           &       0.142   &       0.000    &       0.151   &       -0.027 &       0.140\\
\hline
\hline
\end{tabular}
\end{center}
\end{table}

\clearpage
\newpage

\begin{table}
\begin{center}
\caption{\label{tab:zpe}Energy expectation values (in \cm) of the
different isotopomers in the vibrational ground state. The
subscripts refer to single sets for one set operators or to pairs
of sets for the two-set operators, e.g.. $V_{12}=\langle
V^{(2)}(\bg_{1},\bg_{2})\rangle$. Note that $V_{0}$=462 \cm in
Eq. (\ref{eq:pes}).}
\begin{tabular}{c|cccc|ccccccc|c}
\hline
\hline
& $T_1$ & $T_2$ & $T_3$ & $T$ & $V_{1}$ & {$V_{2}$} & {$V_{3}$} & $V_{12}$ & {$V_{13}$} & {$V_{23}$} &{$V$} & {$T$+$V$} \\
\hline
{HOHOH$^-$} & 1995 & 338 & 935 & 3268 & 1976 & 425  & 944 & -5 & -80 & -384 & 3338 & 6606\\
{HODOH$^-$} & 1997 & 349 & 646 & 2992 & 1964 & 423 & 681  & -5 & -71 & -359   & 3095 & 6087\\
{DOHOH$^-$} & 1741 & 299 & 940 & 2980 & 1731 & 392 & 994   & -4 & -122 & -428  & 3025 & 6005\\
{DODOH$^-$} & 1745 & 309 & 651 & 2705 & 1698 & 374 & 685      & -5 & -68 & -368   & 2778 & 5483\\
{DOHOD$^-$} & 1489 & 259 & 941  &2689 & 1459 & 385 & 1005     & -6 & -93 & -496 & 2716 & 5405\\
{DODOD$^-$} & 1491 & 268 & 653 & 2412 & 1445 & 378 & 740  & -6 & -83 & -467   & 2469 & 4881\\
\hline
\hline
\end{tabular}
\end{center}
\end{table}

\clearpage
\newpage
\begin{table}[htdp]
\caption{Expectation values for the O-O distance (in \AA{}) and
reductions for substitution of BH for the different isotopomers.
The classical value at the transition state is 2.446\AA.}
\begin{center}
\begin{tabular}{|c r || c r || c r |}
\hline
case &$\langle R_{\rm O-O} \rangle $ & case &$\langle R_{\rm O-O} \rangle $ & case & $\langle R_{\rm O-O} \rangle $\\
\hline
HOHOH$^-$ & 2.492 & DOHOH$^-$ &2.495 & DOHOD$^-$ & 2.493\\
HODOH$^-$ & 2.488 & DODOH$^-$ &2.486 & DODOD$^-$ & 2.488\\
\hline
\end{tabular}
\end{center}
\label{tab:oo}
\end{table}%
\clearpage
\newpage

\clearpage
\newpage

\clearpage
\newpage

\end{document}